\documentclass[12pt,twoside,dvips]{article}
\usepackage{epsfig}
\usepackage{epstopdf}
\input{epsf.sty}
\usepackage{epsf}
\usepackage{amssymb}
\usepackage{amsfonts}
\usepackage{amsmath}
\usepackage{pifont}
\usepackage[spanish,english]{babel}
\usepackage{graphicx}
\usepackage{verbatim}

\begin{document}

\title{Constraints on 3-3-1 models with electroweak $Z$ pole observables and $Z'$ search at LHC}
\author{R. Mart\'{\i}nez$\thanks{%
e-mail: remartinezm@unal.edu.co}$, F. Ochoa$\thanks{%
e-mail: faochoap@unal.edu.co}$, \and Departamento de F\'{\i}sica, Universidad
Nacional de Colombia, \\ Bogot\'{a} D.C, Colombia}
 
\maketitle

\begin{abstract}
In the framework of general 3-3-1 models, we perform a global $\chi ^2$ fit at $95\%$ C.L. to the updated Z pole observables, including new data from atomic parity violation. We found scenarios where specific realizations of 3-3-1 models are largely excluded above the current limit on the extra neutral gauge boson ($Z'$) research at LHC. We derive a relation between the $Z-Z'$ mixing angle and the parameter $\tan(\beta)$ of the effective Two-Higgs-Doublet Model contained into the 3-3-1 spectrum. We show that particular versions of 3-3-1 models  containing particles with exotic electric charges are disfavored by the electroweak precision observables in an ample region of the parameter space.

\end{abstract}

\section{Introduction}

After the observation of an 125 GeV scalar particle at CERN-LHC by the ATLAS and CMS collaborations  \cite{scalar_signal_ATLAS,scalar_signal_CMS}, further analysis have continued to test the compatibility of the data with the Standard Model (SM) predictions. Although most of the experimental results are compatible with the SM, there are still some unexplained discrepancies and theoretical issues that the SM leaves unanswered. For example, at LHC, excess signal events in the diphoton channel $h\rightarrow\gamma \gamma$ \cite{diphoton-1,diphoton-2} and 4-lepton channel  $h\rightarrow Z Z^*\rightarrow 4l$ \cite{diz-1,diz-2} have been reported. The BaBar collaboration has found a discrepancy in the value of  $\overline{B} \rightarrow D^{(*)}\tau ^-\overline{\nu_{\tau}}$ \cite{babar}. Also, Tevatron has observed an anomalously large forward-backward asymmetry in top quark pairs \cite{FBtop}. Regarding the theoretical aspects, the SM does not explain the existence of three families and the mass hierarchy \cite{Langacker1,Fayet}, the matter and anti-matter asymmetry \cite{Langacker1}, the neutrino oscillations \cite{fuku}, the origin of small Flavor Changing Neutral Current (FCNC) processes \cite{FCNC-SM},  etc.

Some of the above problems can be understood by introducing a larger particle content or enlarging the group of symmetry of the SM. This is the case of the group $SU(3)_c \otimes SU(3)_L \otimes U(1)_X$, or 3-3-1 models \cite{331, twelve}, which exhibit the following motivations:  First of all, from the cancellation of chiral anomalies \cite{chiral} and asymptotic freedom in QCD, the 3-3-1 models can explain why there are three fermion
families. Secondly, since the third family is treated under a different representation, the
large mass difference between the heaviest quark family and the two lighter ones may be
understood \cite{mass-diff}. Third, these models contain a natural Peccei-Quinn symmetry, necessary to solve the strong-CP problem \cite{strong-cp}. Also, these models allow to predict the quantization of electric charge and the vectorial character of the electromagnetic interactions \cite{charge-quantum}. 

The group structure of these models leads, along with the SM-neutral boson Z, to the prediction of an additional current associated with a new neutral boson $Z'$, which produces low energy deviations through a $Z-Z'$ angle mixing.  Restrictions to these deviations have been studied in \cite{z-pole-1} and \cite{z-pole-2} using experimental values of Z pole observables from the CERN-LEP and the SLAC Linear Collider, and data from atomic parity violation.

On the other hand, the 3-3-1 models extend the scalar sector of the SM into three $SU(3)_L$ scalar triplets: one heavy triplet field with a Vacuum Expectation Value (VEV) at a high energy scale $\langle \chi \rangle =  \upsilon _{\chi}$, which produces the breaking of the symmetry $SU(3)_L \otimes U(1)_X$ to the SM electroweak group $SU(2)_L \otimes U(1)_Y$, and two lighter triplets with VEVs $\langle \rho \rangle = \upsilon _{\rho}$ and $\langle \eta \rangle = \upsilon _{\eta}$, which induce the breakdown at the electroweak scale according to the relation $ \upsilon ^2 =  \upsilon _{\rho} ^2 +  \upsilon _{\eta} ^2$. Further, it has been shown that 3-3-1 models possess specialized Two Higgs Doublet Models (2HDM) \cite{thdm} as effective low energy models, where the light scalar triplets are decomposed into two hypercharge-one $SU(2)_L$ doublets and scalar singlets \cite{higgs-production}.  Thus, in addition to the above properties, the 3-3-1 models heritage all the interesting phenomenology exhibited by 2HDM, as the implementation of additional sources of CP violation and FCNC \cite{thdm}, generation of baryon asymmetry in the Universe \cite{baryon}, connections with a tree-level minimal supersymmetric standard model \cite{susy}, etc. 

In the context of 2HDM, many studies have imposed different constraints on the parameters of the model \cite{2hdm-const-1}, which could be extended to the low energy behavior of 3-3-1 models, including constraints on various couplings of the Higgs potential \cite{potential-const}, the masses of the neutral and charged Higgs bosons, and the parameter  $\tan(\beta)$ defined as the ratio between the VEVs of the two scalar doublets \cite{B-fisica1, B-fisica2, B-fisica3, LEP-const, LHC-const}. In particular, fits on $\tan(\beta)$ is an important issue because the constraints on this parameter allows to find excluded areas from specific realizations of 2HDM. In the 3-3-1 case, the parameter $\tan(\beta)$ is  defined in the same form as in 2HDM, and the constraints on this parameter is applicable to 3-3-1 models. However, due to the $Z-Z'$ mixing, there arises a functional relation between the mixing angle and $\tan(\beta)$, according to Eq. (\ref{mixing-angle-2}) below, which lead us to additional restrictions that involves effects due to the extra $Z'$ boson, which is not found in minimal 2HDMs.

The purpose of this paper is two-fold: First, update some constraints obtained in \cite{z-pole-1} and \cite{z-pole-2} taking into account the most recent calculations of the Z pole observables in the SM, the new data from atomic parity violation, and the current limit on the extra neutral gauge boson research at LHC. We found that the global least-square ($\chi ^2$) fit to the updated data exhibits appreciable variations. Second, in \cite{z-pole-1, z-pole-2} the $Z-Z'$ mixing was taken as a free parameter regardless of its relation with $\tan (\beta)$. Thus, we reconstruct the analysis using the parameter $\tan(\beta)$ instead of the mixing angle, which allows to obtain new constraints on this parameter different from other analysis obtained in the context of minimal 2HDMs.  

In Sec. 2, we show the particle content of the model, specifically, the fermonic, scalar and neutral gauge sectors. In Sec. 3,  we show the weak couplings taking into account the small $Z-Z'$ mixing angle. Sec. 4 is devoted to study the Z-pole constraints from a global $\chi ^2$ fit. Finally, in Sec. 5 we summarize our results. An appendix is added, where we show in detail the analytical expressions and data for the Z pole observables.

\section{Particle content} 

In general 3-3-1 models, the electric charge is defined by the combination

\begin{eqnarray}
Q=T_3+b T_8+X,
\label{electric-charge}
\end{eqnarray}
with $T_3=\frac{1}{2}Diag(1,-1,0)$, $T_8=(\frac{1}{2\sqrt{3}})Diag(1,1,-2)$, $X$ the $U(1)_X$ charge, and $b$ a free parameter which determines different types of 3-3-1 models \footnote{Instead of the parameter  $\beta$ in \cite{z-pole-1, z-pole-2}, we use $b$ in order to avoid confusion with the parameter $\tan(\beta)$.}. In order to avoid chiral anomalies, the model introduces in the fermionic sector the following $(SU(3)_c, SU(3)_L,U(1)_X)$ left- and right-handed representations:

\begin{eqnarray}
Q^{1}_L
&=&
\begin{pmatrix}
U^{1} \\
D^{1} \\
T^{1} \\
\end{pmatrix}_L:\left(3,3,\frac{1-\sqrt{3}b}{6}\right) ,
\left\{
\begin{array}{c}
U^{1}_R :(3^*,1,2/3) \\
D^{1}_R :(3^*,1,-1/3) \\
T^{1}_R :\left(3^*,1,\frac{1-3\sqrt{3}b}{6}\right) \\
\end{array}
\right.  \nonumber \\
Q^{2,3}_L
&=&
\begin{pmatrix}
D^{2,3} \\
U^{2,3} \\
J^{2,3} \\
\end{pmatrix}_L:(3,3^*,\frac{1+\sqrt{3}b}{6}), \left\{
\begin{array}{c}
D^{2,3}_R :(3^*,1,-1/3) \\
U^{2,3}_R :(3^*,1,2/3) \\
J^{2,3}_R :\left(3^*,1,\frac{1+3\sqrt{3}b}{6}\right) \\
\end{array}
\right. \nonumber  \\
L^{1,2,3}_L
&=&
\begin{pmatrix}
\nu ^{1,2,3} \\
e^{1,2,3} \\
E^{1,2,3} \\
\end{pmatrix}_L :\left(1,3,\frac{-\sqrt{3}-b}{2\sqrt{3}}\right) ,\left\{
\begin{array}{c}
e^{1,2,3}_R :(1,1,-1) \\
E_R^{1,2,3}:\left(1,1,\frac{-1-\sqrt{3}b}{2}\right) \\
\end{array}
\right.,
\label{fermion_spectrum}
\end{eqnarray}
where $U^{i}$ and $D^{i}$ for $i=1,2,3$ are three up- and down-type quark components in the flavor basis,  while $\nu ^{i}$ and $e^{i}$ are the neutral and charged lepton families. The right-handed sector transforms as singlets under $SU(3)_L$ with $U(1)_X$ quantum numbers equal to the electric charges. In addition, we see that the model introduces extra fermions with the following properties: a single flavor quark  $T^{1}$ with electric charge $(1-3\sqrt{3}b)/6$, two flavor quarks $J^{2,3}$ with charge $(1+3\sqrt{3}b)/6$, and three leptons $E^{1,2,3}$ with charge $(-1-\sqrt{3}b)/2$. The most popular 3-3-1 models correspond to $b= \sqrt{3}$ ($-\sqrt{3}$) \cite{331}, where the extra quark spectrum has two quarks with exotic electric charge of $5/3$ ($-4/3$) and one with charge $-4/3$ ($5/3$), and models with $b= 1/\sqrt{3}$ ($b= -1/\sqrt{3}$) \cite{twelve} that have two extra quarks with ordinary electric charge of $2/3$ ($-1/3$) and one with charge $-1/3$ ($2/3$). In particular, models with $b=\sqrt{3}$ posses in addition to the exotic quarks, leptons with charge $-2$ (we call them exotic models), while the model with $b=-\sqrt{3}$ have extra leptons with ordinary charge $1$ (we call them quark-exotic models). In contrast, models with $b= 1/\sqrt{3}$ have extra leptons with charge $-1$ (non-exotic models), while  $b= -1/\sqrt{3}$ posses extra neutral leptons (neutral non-exotic models). In Tab. \ref{popular}, we resume the above assignation for the four typical 3-3-1 models.       

We emphasize that the chosen representation in  (\ref{fermion_spectrum}) is not unique. There is another alternative, corresponding to the conjugate representation of (\ref{fermion_spectrum}), which is considered by authors in \cite{buras, buras-2}. However, both schemes are equivalent if we change the sign of the parameter $b$ in Eq. (\ref{electric-charge}). Thus, in this scheme, the exotic model corresponds to $b=-\sqrt{3}$, the quark-exotic model is $b=\sqrt{3}$, the non-exotic model is $b=-1/\sqrt{3}$ and the neutral non-exotic model is $b=1/\sqrt{3}$.  

On the other hand, the scalar sector introduces one triplet field with VEV $\langle \chi \rangle_0=\upsilon_{\chi}$, which provides the masses to the new heavy fermions, and two triplets with VEVs $\langle \rho \rangle_0=\upsilon_{\rho}$ and $\langle \eta \rangle_0=\upsilon_{\eta}$, which give masses to the SM fermions at the electroweak scale. The $(SU(3)_L,U(1)_X)$ group structure of the scalar fields are

\begin{eqnarray}
\chi&=&
\begin{pmatrix}
\chi_1\\
\chi_2 \\
\frac{1}{\sqrt{2}}(\upsilon_{\chi} + \xi _{\chi} + i \zeta_{\chi} ) \\
\end{pmatrix}: \left(3,\frac{b}{\sqrt{3}}\right) \notag \\
\rho&=&
\begin{pmatrix}
\rho_1^+ \\
\frac{1}{\sqrt{2}}(\upsilon_{\rho} + \xi _{\rho} + i \zeta_{\rho} ) \\
\rho _3 \\
\end{pmatrix} : \left(3,\frac{\sqrt{3}-b}{2\sqrt{3}}\right) \notag \\
\eta &=&
\begin{pmatrix}
\frac{1}{\sqrt{2}}(\upsilon_{\eta} + \xi _{\eta} + i \zeta_{\eta} ) \\
\eta _2^{-} \\
\eta _3
\end{pmatrix}:\left(3,\frac{-\sqrt{3}-b}{2\sqrt{3}}\right).
\label{331-scalar}
\end{eqnarray}

We will use the following parametrizations on the VEVs of $\rho$ and $\eta $ \cite{buras-2}:

\begin{eqnarray}
\upsilon _{\pm}^2&=&\upsilon _{\eta}^2\pm \upsilon _{\rho}^2, \nonumber \\
\tan(\beta)&=&T_{\beta}=\frac{\upsilon _{\rho}}{\upsilon _{\eta}} .
\label{tanbeta}
\end{eqnarray}

For the boson vector spectrum, we are just interested in the physical neutral sector that corresponds to the photon $A$, the neutral weak boson $Z$ and a new neutral boson $Z'$,  which are written in terms of the electroweak $SU(3)_L \otimes U(1)_X$ gauge fields as:

\begin{eqnarray}
A_{\mu } &=&S_{W}W_{\mu }^{3}+C_{W}\left( b T_{W}W_{\mu }^{8}+\sqrt{%
1-b ^{2}T_{W}^{2}}B_{\mu }\right) ,  \notag \\
Z_{\mu } &=&C_{W}W_{\mu }^{3}-S_{W}\left( b T_{W}W_{\mu }^{8}+\sqrt{%
1-b ^{2}T_{W}^{2}}B_{\mu }\right) ,  \notag \\
Z_{\mu }^{\prime } &=&-\sqrt{1-b ^{2}T_{W}^{2}}W_{\mu }^{8}+b
T_{W}B_{\mu },
\end{eqnarray}%
where the Weinberg angle is defined as 
\begin{equation}
\sin \theta _{W}=S_{W}=\frac{g^{\prime }}{\sqrt{g^{2}+\left( 1+b
^{2}\right) g^{\prime 2}}},
\label{weinberg}
\end{equation}%
and $g,$ $g^{\prime }$ correspond to the coupling constants of the  $%
SU(3)_{L}$ and $U(1)_{X}$ groups, respectively. However, there is still a small mixing in the basis $(Z,Z')$ given by the symmetrical mixing matrix

\begin{eqnarray}
M^2_{NC}=\begin{pmatrix}
M_{Z}^2 && M_{ZZ'}^2\\
* && M_{Z'}^2\\
\end{pmatrix},
\end{eqnarray}
where

\begin{eqnarray}
M_{Z}^{2}&\simeq &\frac{g^{2}}{4C_{W}^{2}} \upsilon _{+ }^{2}, \nonumber \\
M_{Z'}^{2}&\simeq & \frac{g'^{2}}{3T_{W}^{2}}\upsilon _{\chi }^{2}, \nonumber \\
M_{ZZ'}^2&=&\frac{-gg'}{12C_{W}}\left[ 3b T_{W}\upsilon _{+}^{2}+ \sqrt{3}T_{W}^{-1}\upsilon _{-}^{2} \right],
\label{masas-gauge-neutro}
\end{eqnarray}%
which diagonalize to the true mass eigenstates through the rotation

\begin{eqnarray}
\begin{pmatrix}
Z_{\mu}\\
Z'_{\mu}\\
\end{pmatrix}
=R_{\theta}\begin{pmatrix}
Z_{1\mu } \\
Z_{2\mu }  \\
\end{pmatrix}
\label{neutral-rotation}
\end{eqnarray}
with
\begin{eqnarray}
R_{\theta}=\begin{pmatrix}
C_{\theta} && -S_{\theta}\\
S_{\theta} && C_{\theta}\\
\end{pmatrix}.
\label{rotation}
\end{eqnarray}

The mixing angle in (\ref{rotation}) accomplish the relation \cite{buras-2}:
\begin{eqnarray}
\tan(2\theta)&=&\frac{gg^{\prime }T_{W}^{2}\left[ 3bS_{W}^{2} \upsilon _{+ }^{2} +\sqrt{3}C_{W}^{2} \upsilon _{-}^{2} \right] }{2S_{W}C_{W}^{2}g^{\prime 2}\upsilon _{\chi }^{2}-\frac{3}{2}S_{W}T_{W}^{2}g^{2}\upsilon _{+}^{2} }.
\label{mixing-angle-1}
\end{eqnarray}

We emphasize that the Eq. (\ref{mixing-angle-1}) corresponds to the inverse $\Lambda ^{-1}$ in Eq. (5) of \cite{z-pole-1} but differs in the following aspects: first, the parameter $\beta$ is changed by $b$. Second, in \cite{z-pole-1}, there is a $\sqrt{3}$ factor missing in the denominator. Also, they are different in a global sign. The correct expression was obtained by authors in \cite{buras-2}. Taking into account the definitions from (\ref{tanbeta}), (\ref{weinberg}), and of $M_Z^2$ and $M_{Z'}^2$ in (\ref{masas-gauge-neutro}), the mixing angle can be written as:

\begin{eqnarray}
S_{\theta} &\simeq &\frac{1}{2}\tan(2\theta)\simeq f[b,T_{\beta}]\left(\frac{M_Z}{M_{Z'}}\right)^2, \nonumber \\
f[b,T_{\beta}]&=&\frac{C_W}{\sqrt{1-b^2T_W^2}}\left[bT_W^2+\frac{1}{\sqrt{3}}\left(\frac{1-T_{\beta}^2}{1+T_{\beta}^2}\right)\right],
\label{mixing-angle-2}
\end{eqnarray}
where we have taken the dominant contribution of the Eq. (\ref{mixing-angle-1}) assuming that $\upsilon_{+} \ll \upsilon_{\chi}$.
 
\section{Neutral weak couplings }

With the particle content from Eq. (\ref{fermion_spectrum}) and taking into account the rotations to mass eigenstates from (\ref{neutral-rotation}), the neutral weak Lagrangian can be written as \cite{z-pole-1}:

\begin{equation}
\mathcal{L}_{D}^{NC}=\frac{g}{2C_{W}}\left[ \overline{f^i}\gamma _{\mu
}\left( G_{v}^{i}-G_{a}^{i}\gamma _{5}\right) f^iZ^{\mu }_1+\overline{f^i}\gamma
_{\mu }\left( \widetilde{G}_{v}^{i}-\widetilde{G}_{a}^{i}\gamma _{5}\right)
f^iZ^{\mu }_2\right] ,  \label{L-neutro}
\end{equation}
where $f^i$ runs over all individual fermions ($f^i=U^i, D^i, \nu ^i, T^1, etc.$), $G_{v,a}^{i}$ are the vector and axial weak couplings to $Z^{\mu}_1$, and  $\widetilde{G}_{v,a}^{i}$ the corresponding weak couplings to $Z^{\mu}_2$. A sum over the indices $i$ is understood. The weak couplings are defined by:

\begin{eqnarray}
G_{v,a}^i&=&g_{v,a}^i+\widetilde{g}_{v,a}^{i}S_{\theta}, \nonumber \\
\widetilde{G}_{v,a}^i&=&\widetilde{g}_{v,a}^i+g_{v,a}^{i}S_{\theta}, 
\end{eqnarray} 
where $g_{v,a}^i$ and $\widetilde{g}_{v,a}^{i}$ are shown in Tab. \ref{EW-couplings}, and $S_{\theta}$ is the mixing angle given by Eq. (\ref{mixing-angle-2}). The usual SM weak neutral currents are obtained in the limit of $S_{\theta}=0$, where the $Z_1$ becomes the SM weak boson $Z$.

As a result of the mixing in the coupling with $Z_1$, small deviations of the SM observables arise. In particular, we use the Z pole observables shown in Tab. \ref{tab:observables} with their experimental values from CERN
collider (LEP), SLAC Liner Collider (SLC) and data from atomic parity
violation \cite{one}, the SM predictions, and the expressions predicted by
3-3-1 models. The deviations due to the 3-3-1 mixing were obtained in \cite{z-pole-2}, which we resume in Appendix \ref{app:z-pole}. We use $M_{Z}=91.1876$ $GeV$, $S_{W}^{2}=0.2310$, and for
the predicted SM partial decay we use the
values from Eq. (\ref{SM-partial-decay}). For the quark sector in (\ref{fermion_spectrum}), we use the following assignation to the phenomenological quarks:
 
 \begin{eqnarray}
 U^{1,2,3}=(t,u,c), \ \ \  D^{1,2,3}=(b,d,s),
\label{quark-assign} 
 \end{eqnarray}
such that the $(t,b)$ quarks transform as the conjugate of $(u,d)$ and $(c,s)$.

\section{Z pole constraints}

With the expressions for the Z pole observables and the experimental data
shown in table \ref{tab:observables}, we perform a $\chi ^{2}$ fit at 95\% CL, where the free quantities $%
T_{\beta },$ $M_{Z'}$ and $b $ can be constrained at the $Z$
peak. We assume a covariance matrix with elements $V_{ij}=\rho _{ij}\sigma
_{i}\sigma _{j}$ among the Z pole observables, where $\rho $ is the correlation
matrix and $\sigma $ the quadratic root between the experimental and SM errors.
The $\chi ^{2}$ statistic with three degrees of freedom (d.o.f) is defined
as \cite{one}

\begin{equation}
\chi ^{2}(T_{\beta },M_{Z'},b )=\left[ \mathbf{y}-\mathbf{F}%
(T_{\beta },M_{Z'},b )\right] ^{T}V^{-1}\left[ \mathbf{y}-\mathbf{F}%
(T_{\beta },M_{Z'},b )\right] ,  \label{chi}
\end{equation}
where $\mathbf{y=\{}y_{i}\mathbf{\}}$ represent the 22
experimental observables from table \ref{tab:observables}, and $\mathbf{F}$
the corresponding 3-3-1 prediction. Table \ref{tab:correlation} from appendix %
\ref{app:z-pole} display the symmetrical correlation matrices taken from
ref. \cite{LEP}. The function in (\ref{chi}) imposes constraints to the free variables ($T_{\beta
},M_{Z'},b $)  by requiring that $\chi ^{2}\leq \chi _{\min}^{2}+\chi _{CL}^{2}$.
For three d.o.f at $95\%$ C.L., we have $\chi _{CL}^{2}=7.815$. For the assignation in (\ref{quark-assign}) and the Z pole parameters, we find for the minimum that $\chi _{\min }^{2}=160.24$. 

\begin{figure}[t]
\centering \includegraphics[scale=0.8]{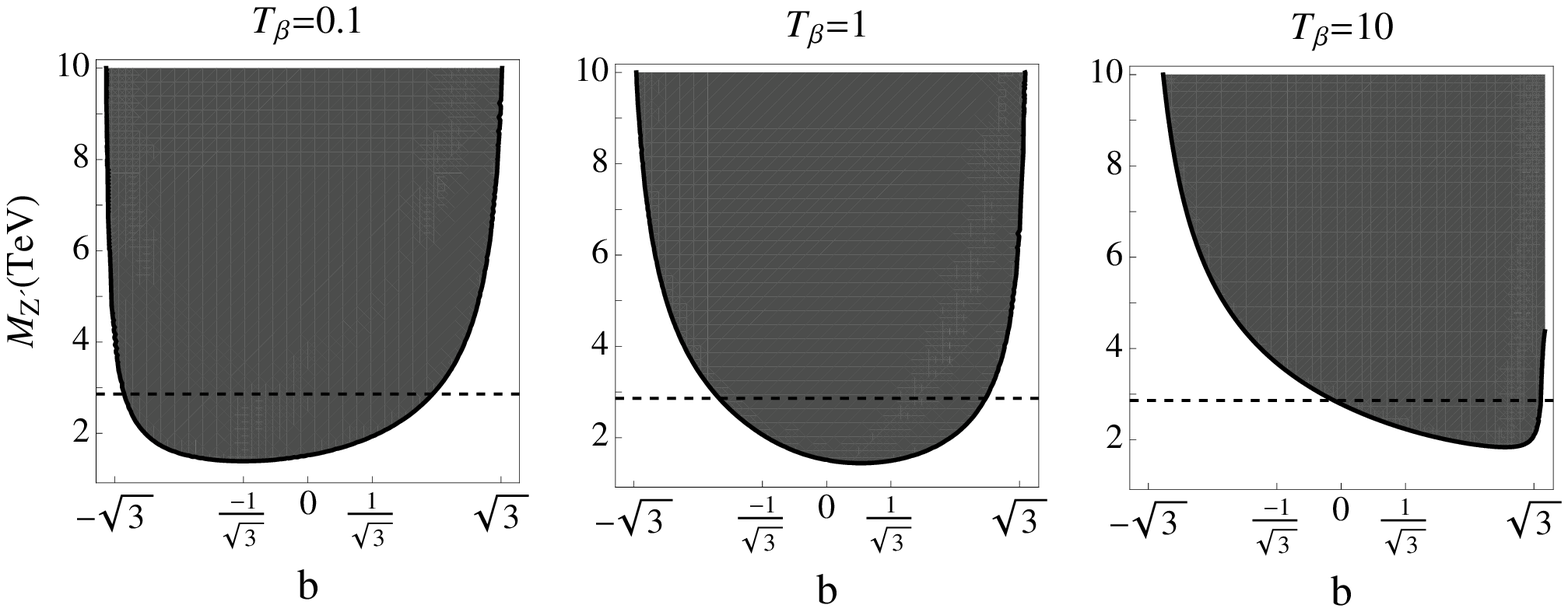}
\caption{Constraints for $M_{Z'}$ in the $(M_{Z'},b)$ plane for three values of $T_{\beta}$. The shaded areas show the allowed points from Z pole constraints. The dashed line is the experimental lowest limit $M_{Z'}=2.86$ TeV from LHC.}
\label{fig-1}
\end{figure}

We obtain 2-dimensional projections of the $\chi ^{2}$ function with the above restrictions. Figure \ref{fig-1} shows the limits for the $Z'$ mass in the $(M_{Z'},b)$ plane for three values of  $T_{\beta}$: $0.1, 1$ and $10$. The dashed horizontal lines mark the lowest limit $M_{Z'}=2.86$ TeV for extra neutral gauge boson research at LHC \cite{ZP-limits}. If we consider the four typical 3-3-1 models, we observe that:

\begin{enumerate}
 
\item In the scenario with $T_{\beta}=0.1$, we see that the quark-exotic model ($b=-\sqrt{3} $) is excluded for $M_Z'< 3$ TeV, while the exotic model ($b=\sqrt{3} $) is rejected for $M_{Z'}<7$ TeV. The models with $b=\pm 1/\sqrt{3} $ have allowed points in all the range above the $Z'$ experimental limit.
 
\item In the scenario with $T_{\beta}=1$, the quark-exotic model is excluded for $M_{Z'} < 8.5$ TeV, while for the exotic model the exclusion happens for $M_{Z'} < 5$ TeV. Again, the other two models show allowed points from the experimental limit.
 
\item In the scenario with $T_{\beta}=10$, the quark-exotic model is largely excluded, without limits in the range to 10 TeV. In addition, we see that this scenario shows that the neutral non-exotic model ($b=-1/\sqrt{3}$) is excluded for $M_{Z'} < 4$ TeV.
 
 \end{enumerate}
We found that the constraints does not change significantly for values above $T_{\beta}=10$.
\\

\begin{figure}[b]
\centering\includegraphics[scale=0.8]{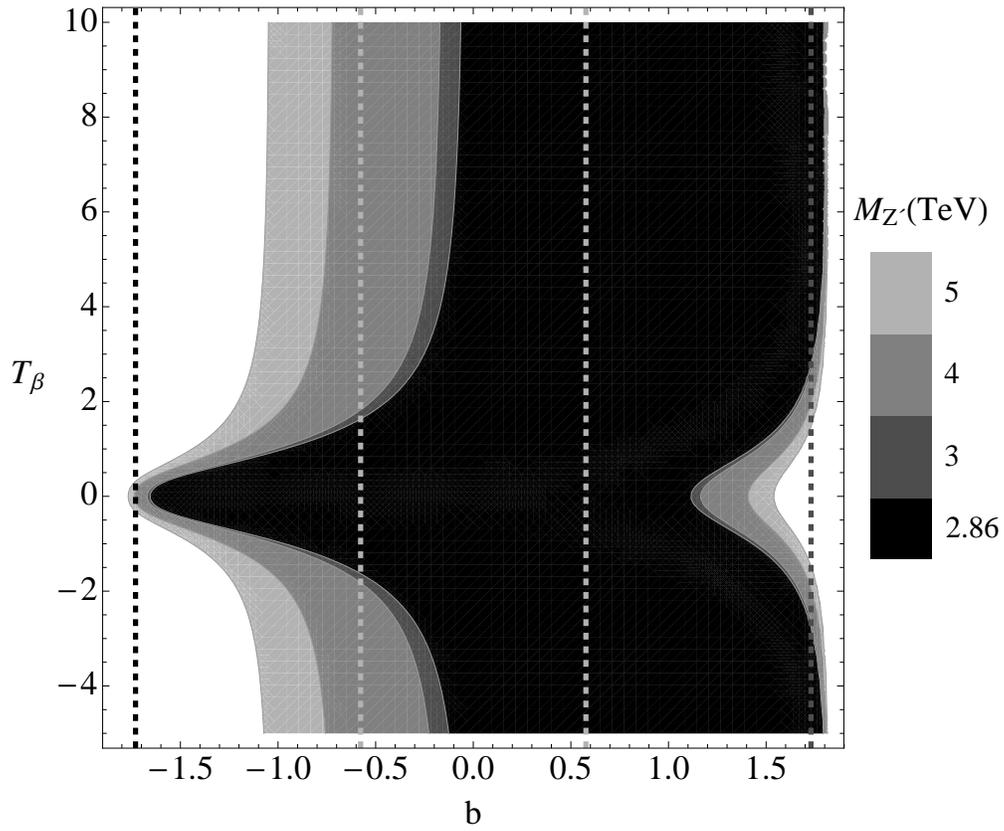}
\caption{Constraints for $T_{\beta}$ in the $(T_{\beta},b)$ plane. The central black area shows the allowed area for the experimental limit $M_{Z'}=2.86$ TeV from LHC. The shaded areas show the limits for three values of $M_{Z'}$. The dashed lines mark the models with $b=-\sqrt{3}, -1/\sqrt{3}, 1/\sqrt{3}$ and $\sqrt{3}$, respectively.}
\label{fig-2}
\end{figure}

On the other hand, figure \ref{fig-2} shows the constraints for $T_{\beta}$ in the $(T_{\beta},b)$ plane. The central black area shows the allowed points for the experimental limit of $M_{Z'}=2.86$ TeV. The shaded areas show the allowed limits for three values of $M_{Z'}$: 3, 4, and 5 TeV. The dashed vertical lines mark the four 3-3-1 models with $b=-\sqrt{3}, -1/\sqrt{3}, 1/\sqrt{3}$ and $\sqrt{3}$, respectively. We see that:

\begin{enumerate}

\item  The non-exotic model ($b=1/\sqrt{3}$) is allowed for all value of $T_{\beta}$ and $M_{Z'}$ above 2.86 TeV. 

\item The neutral non-exotic model excludes the area with $T_{\beta}> \vert1.5\vert$ for $M_{Z'}=2.86$ TeV and $T_{\beta}> \vert1.7\vert$ for $M_{Z'}=3$ TeV. There is not exclusion bounds for $M_{Z'}=4$ and $5$ TeV.

\item In contrast to the above model, the exotic model excludes a central band around zero. The excluded areas are: $T_{\beta}< \vert2.5\vert$ for $M_{Z'}=2.86$ and $3$ TeV, $T_{\beta}< \vert2\vert$ for $M_{Z'}=4$ TeV, and $T_{\beta}< \vert1.5\vert$ for $M_{Z'}=5$ TeV.

\item The quark-exotic model is completely excluded for $M_{Z'}=2.86$ and $3$ TeV. Only, one point at $T_{\beta}=0$ is allowed for $M_{Z'}=4$ TeV, and a small range $T_{\beta}< \vert0.25\vert$ is allowed for $M_{Z'}=5$ TeV.

\end{enumerate}

\begin{figure}[b]
 \centering\includegraphics[scale=0.75]{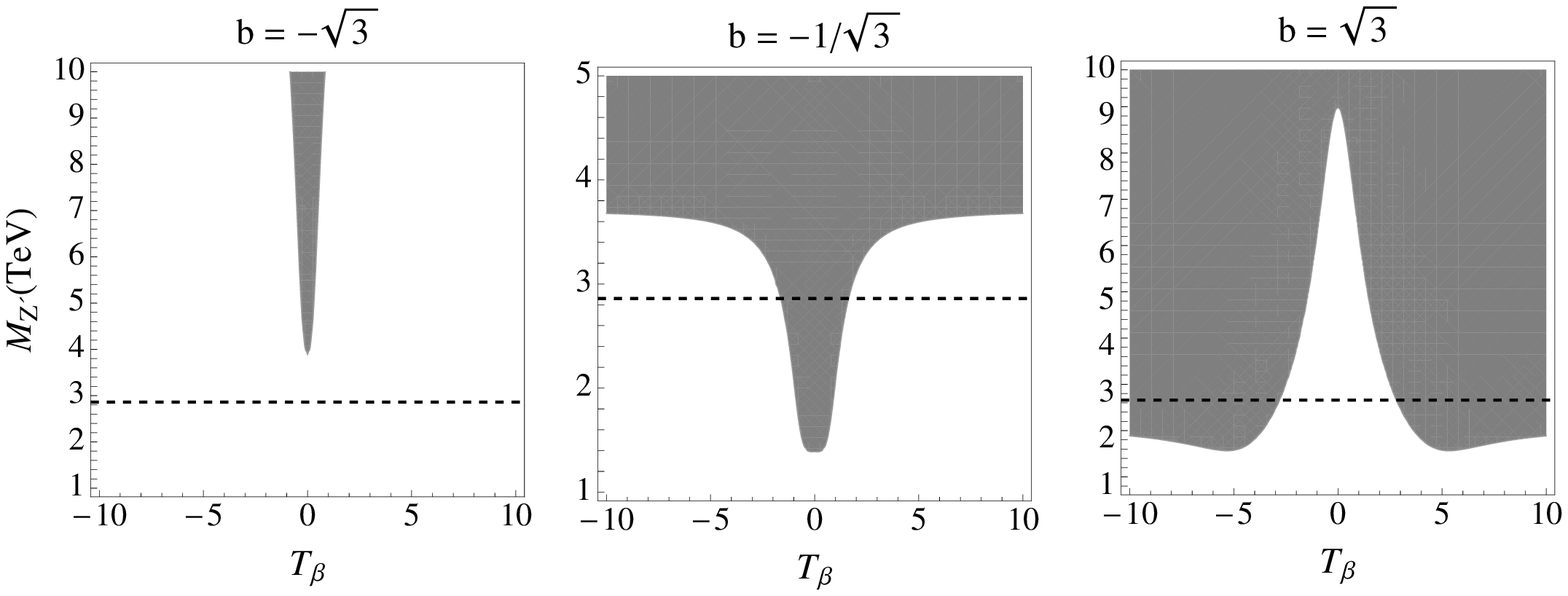}
\caption{Constraints in the $(M_{Z'},T_{\beta})$ plane for three values of $b$. The shaded areas show the allowed points from Z pole constraints. The dashed line is the experimental lowest limit $M_{Z'}=2.86$ TeV from LHC.}
\label{fig-3}
\end{figure}

Finally, we display the plots in the $(M_{Z'},T_{\beta})$ plane for $b=-\sqrt{3},-1/\sqrt{3}$ and $\sqrt{3}$, as shown in Fig. \ref{fig-3}. We do not show this projection for the non-exotic model since it does not exhibit any excluded area for $M_{Z'}$ above $2.86$ TeV. We obtain that:

\begin{enumerate}

\item The quark-exotic model is excluded for all values of $T_{\beta}$ if $M_{Z'}<4$ TeV. Above 4 TeV, this model exhibits allowed points in small ranges of $\tan{\beta}$.

\item The neutral non-exotic model exhibits an allowed central region around zero, which increases with $M_{Z'}$. Out from this bounds, the model is excluded. However, for $M_{Z'} \gtrsim 3.7$ TeV, the model is allowed for all $T_{\beta}$ range without limits.

\item For the exotic model, a central region is excluded. However, this exclusion bounds decreases with $M_{Z'}$ and disappears at $M_{Z'} \approx 9.1$ TeV. 
 
 \end{enumerate}

\section{Conclusions}
Exclusion limits for 3-3-1 models were obtained by the electroweak Z pole precision observables at $95\%$ C.L., taking into account the updated SM calculations, and experimental limits from atomic parity violation and  search for $Z'$ resonances at LHC. We conclude that:

\begin{itemize}

\item For $M_{Z'}$ from the experimental lower limit 2.86 TeV to 4 TeV, the quark-exotic 3-3-1-version is completely excluded. Even for masses above 4 TeV, this model is largely constrained, exhibiting allowed points at small $\tan{\beta}$ values, as shown in the left plot of Fig. \ref{fig-3}. 

\item For the exotic model, there is not general exclusion area for values $M_{Z'}\geq 9.1$ TeV. Below this mass, the model is excluded for small values of $\tan{\beta}$. In particular, for the lower experimental limit $M_{Z'}= 2.86$ TeV, this model is excluded in $\tan{\beta}\leq \vert3\vert$.

\item Above $M_{Z'}> 2.86$ TeV, the non-exotic model is compatible with the electroweak Z pole observables for all $\tan{\beta}$ values.

\item The neutral non-exotic version shows exclusion points for $M_{Z'}\leq 3.7$ TeV. At $M_{Z'}= 2.86$ TeV, the model is excluded for $\tan{\beta} > \vert1.7\vert$.

\end{itemize}

\section*{Appendix}

\appendix

\section{Z Pole observables\label{app:z-pole}}

The Z pole parameters with their experimental values from CERN
collider (LEP), SLAC Liner Collider (SLC) and data from atomic parity
violation taken from ref. \cite{one}, are shown in table \ref%
{tab:observables}, with the SM predictions and the expressions predicted by
331 models. The corresponding correlation matrix from ref. \cite{LEP} is
given in table \ref{tab:correlation}.  In the SM, the partial decay widths of $Z_{1}$ into
fermions $f\overline{f}$ is described by \cite{one, pitch}:

\begin{equation}
\Gamma _{f^i}^{SM}=\frac{N_{c}^{f}G_{f}M_{Z_{1}}^{3}}{6\sqrt{2}\pi }\rho _{f}%
\sqrt{1-\mu _{f}^{2}}\left[ \left( 1+\frac{\mu _{f}^{2}}{2}\right) \left(
g_{v}^{i}\right) ^{2}+\left( 1-\mu _{f}^{2}\right) \left( g_{a}^{i}\right)
^{2}\right] R_{QED}R_{QCD},  \label{partial-decay}
\end{equation}
where $N_{c}^{f}=1$, 3 for leptons and quarks, respectively. $%
R_{QED}=1+\delta _{QED}^{f}$ and $R_{QCD}=1+\frac{1}{2}\left(
N_{c}^{f}-1\right) \delta _{QCD}^{f}$ are QED and QCD corrections , and $\mu
_{f}^{2}=4m_{f}^{2}/M_{Z}^{2}$ considers kinematical corrections only
important for the $b$-quark. Universal electroweak corrections sensitive to
the top quark mass are taken into account in $\rho _{f}=1+\rho _{t}$ and in $%
g_{v}^{i}$ which is written in terms of an effective Weinberg angle \cite%
{one}

\begin{equation}
\overline{S_{W}}^{2}=\left( 1+\frac{\rho _{t}}{T_{W}^{2}}\right) S_{W}^{2},
\label{effective-angle}
\end{equation}
with $\rho _{t}=3G_{f}m_{t}^{2}/8\sqrt{2}\pi ^{2}$. Nonuniversal
vertex corrections are also taken into account in the $Z_{1}\overline{b}b$
vertex with additional one-loop leading terms which leads to $\rho _{b}=1-%
\frac{1}{3}\rho _{t}$ and $\overline{S_{W}}^{2}=\left( 1+\frac{\rho _{t}}{%
T_{W}^{2}}+\frac{2\rho _{t}}{3}\right) S_{W}^{2}$ \cite{one, pitch}. 

For the top and bottom quark masses,
we use the following values calculated at the Z pole scale \cite{run-mass}:

\begin{eqnarray}
m_{t}(M_{Z}) &=&171.684 \  GeV, \nonumber \\
m_{b}(M_{Z})&=&2.853 \ GeV.  \label{quarks-mass}
\end{eqnarray}

For the partial SM partial decay given by Eq. (\ref{partial-decay}), we use
the following values taken from ref. \cite{one}

\begin{eqnarray}
\Gamma _{u}^{SM} &=&0.30026\pm 0.00005\text{ }GeV;\quad \Gamma
_{d}^{SM}=0.38304\pm 0.0005\text{ }GeV;  \notag \\
\Gamma _{b}^{SM} &=&0.37598\pm 0.00003\text{ }GeV;\quad \Gamma _{\nu
}^{SM}=0.16722\pm 0.00001\text{ }GeV;  \notag \\
\Gamma _{e}^{SM} &=&0.08400\pm 0.00001\text{ }GeV.  \label{SM-partial-decay}
\end{eqnarray}

For the 3-3-1 model in the fourth column of Tab. \ref{tab:observables}, the analytical deviations are:

\begin{eqnarray}
\delta _{Z} &=&\frac{\Gamma _{u}^{SM}}{\Gamma _{Z}^{SM}}(\delta _{u}+\delta
_{c})+\frac{\Gamma _{d}^{SM}}{\Gamma _{Z}^{SM}}(\delta _{d}+\delta _{s})+%
\frac{\Gamma _{b}^{SM}}{\Gamma _{Z}^{SM}}\delta _{b}+3\frac{\Gamma _{\nu
}^{SM}}{\Gamma _{Z}^{SM}}\delta _{\nu }+3\frac{\Gamma _{e}^{SM}}{\Gamma
_{Z}^{SM}}\delta _{\ell };  \notag \\
\delta _{had} &=&R_{c}^{SM}(\delta _{u}+\delta _{c})+R_{b}^{SM}\delta _{b}+%
\frac{\Gamma _{d}^{SM}}{\Gamma _{had}^{SM}}(\delta _{d}+\delta _{s});  \notag
\\
\delta _{\sigma } &=&\delta _{had}+\delta _{\ell }-2\delta _{Z};  \notag \\
\delta A_{f} &=&\frac{\delta g_{V}^{i}}{g_{V}^{i}}+\frac{\delta 
g_{A}^{i}}{g_{A}^{i}}-\delta _{f},  \label{shift1}
\end{eqnarray}

\noindent where for the light fermions

\begin{equation}
\delta _{f}=\frac{2g_{v}^{i}\delta g_{v}^{i}+2g_{a}^{i}\delta 
g_{a}^{i}}{\left( g_{v}^{i}\right) ^{2}+\left( g_{a}^{i}\right)
^{2}},  \label{shift2}
\end{equation}

\noindent while for the $b$-quark

\begin{equation}
\delta _{b}=\frac{\left( 3-\beta _{K}^{2}\right) g_{v}^{b}\delta g%
_{v}^{b}+2\beta _{K}^{2}g_{a}^{b}\delta g_{a}^{b}}{\left( \frac{%
3-\beta _{K}^{2}}{2}\right) \left( g_{v}^{b}\right) ^{2}+\beta
_{K}^{2}\left( g_{a}^{b}\right) ^{2}}.  \label{shift3}
\end{equation}
where $\beta _K=\sqrt{1-(2m_b/M_Z)^2}$. The above expressions are evaluated in terms of the
effective Weinberg angle from Eq. (\ref{effective-angle}).

The weak charge is written as

\begin{equation}
Q_{W}=Q_{W}^{SM}+\Delta Q_{W}=Q_{W}^{SM}\left( 1+\delta Q_{W}\right) ,
\label{weak}
\end{equation}%
where $\delta Q_{W}=\frac{\Delta Q_{W}}{Q_{W}^{SM}}$. The deviation $\Delta
Q_{W}$ is \cite{cesio} 
\begin{equation}
\Delta Q_{W}=\left[ \left( 1+4\frac{S_{W}^{4}}{1-2S_{W}^{2}}\right) Z-N%
\right] \Delta \rho _{M}+\Delta Q_{W}^{\prime },  \label{dev}
\end{equation}%
and $\Delta Q_{W}^{\prime }$ which contains new physics gives

\begin{eqnarray}
\Delta Q_{W}^{\prime } &=&-16\left[ \left( 2Z+N\right) \left( g_{a}^{e}%
\overset{\sim }{g}_{v}^{u}+\overset{\sim }{g}_{a}^{e}g_{v}^{u}%
\right) +\left( Z+2N\right) \left( g_{a}^{e}\overset{\sim }{g}%
_{v}^{d}+\overset{\sim }{g}_{a}^{e}g_{v}^{d}\right) \right] S_{\theta } 
\notag \\
&&-16\left[ \left( 2Z+N\right) \overset{\sim }{g}_{a}^{e}\overset{\sim }{%
g}_{v}^{u}+\left( Z+2N\right) \overset{\sim }{g}_{a}^{e}\overset{%
\sim }{g}_{v}^{d}\right] \frac{M_{Z_{1}}^{2}}{M_{Z_{2}}^{2}}.
\label{new}
\end{eqnarray}

For cesium, and for the first term in (\ref{dev}) we take the value $\left[ \left( 1+4%
\frac{S_{W}^{4}}{1-2S_{W}^{2}}\right) Z-N\right] \Delta \rho _{M}\simeq
-0.01 $ \cite{cesio2,cesio}.

\section*{Acknowledgments}

This work was supported by the Departamento Administrativo de Ciencia, Tecnolog\'{\i}a e Innovaci\'on (COLCIENCIAS) in Colombia. We are grateful to Andrzej J. Buras from Munich, Tech. U. for very useful discussion and suggestions that motivated this work.

\newpage

\begin{table}
\begin{center}
\begin{tabular}{c|cccc}
Charges &\begin{tabular}{c} Exotic \\ ($\sqrt{3}$) \end{tabular} & \begin{tabular}{c} Quark-exotic \\ ($-\sqrt{3}$) \end{tabular}& \begin{tabular}{c}Non-exotic \\ ($1/\sqrt{3}$)\end{tabular} & \begin{tabular}{c} Neutral non-exotic \\ ($-1/\sqrt{3}$)\end{tabular} \\ \hline \hline
5/3&2&1&0&0 \\
-4/3&1&2&0&0\\
2/3&0&0&2&1 \\
-1/3&0&0&1&2  \\
-2&3&0&0&0 \\
1&0&3&0&0 \\
-1&0&0&3&0 \\ 
0&0&0&0&3 \\ \hline
\end{tabular}
\end{center}
\caption{Number of extra new fermions in the four most popular 3-3-1 models according to its electric charge}
\label{popular}
\end{table}

\begin{table}[h]
\begin{center}
\begin{tabular}{ccccc}
\hline
$Fermion$ & $g_{v}^{i}$ & $g_{a}^{i}$ & $\widetilde{g}_{v}^{i}$ & $%
\widetilde{g}_{a}^{i}$ \\ \hline\hline
$\nu ^{i}$ & $\frac{1}{2}$ & $\frac{1}{2}$ & $\frac{-1+(1-\sqrt{3}b
)S_{W}^{2}}{2\sqrt{3}\sqrt{1-(1+b ^{2})S_{W}^{2}}}$ & $\frac{-1+(1-\sqrt{%
3}b )S_{W}^{2}}{2\sqrt{3}\sqrt{1-(1+b ^{2})S_{W}^{2}}}$ \\ \hline
$e^{i}$ & $-\frac{1}{2}+2S_{W}^{2}$ & $-\frac{1}{2}$ & $\frac{-1+(1-3\sqrt{3}%
b )S_{W}^{2}}{2\sqrt{3}\sqrt{1-(1+b ^{2})S_{W}^{2}}}$ & $\frac{-1+(1+%
\sqrt{3}b )S_{W}^{2}}{2\sqrt{3}\sqrt{1-(1+b ^{2})S_{W}^{2}}}$ \\ 
\hline
$E^{i}$ & $2Q_{E}S_{W}^{2}$ & $0$ & $\frac{1-(1+2\sqrt{3}b
Q_{E})S_{W}^{2}}{\sqrt{3}\sqrt{1-(1+b ^{2})S_{W}^{2}}}$ & $\frac{%
1-S_{W}^{2}}{\sqrt{3}\sqrt{1-(1+b ^{2})S_{W}^{2}}}$ \\ \hline
$D^{n}$ & $-\frac{1}{2}+\frac{2}{3}S_{W}^{2}$ & $-\frac{1}{2}$ & $%
\frac{3-(3+\sqrt{3}b )S_{W}^{2}}{6\sqrt{3}\sqrt{1-(1+b ^{2})S_{W}^{2}%
}}$ & $\frac{1-(1-\sqrt{3}b )S_{W}^{2}}{2\sqrt{3}\sqrt{1-(1+b
^{2})S_{W}^{2}}}$ \\ \hline
$U^{n}$ & $\frac{1}{2}-\frac{4}{3}S_{W}^{2}$ & $\frac{1}{2}$ & $%
\frac{3-(3-5\sqrt{3}b )S_{W}^{2}}{6\sqrt{3}\sqrt{1-(1+b
^{2})S_{W}^{2}}}$ & $\frac{1-(1+\sqrt{3}b )S_{W}^{2}}{2\sqrt{3}\sqrt{%
1-(1+b ^{2})S_{W}^{2}}}$ \\ \hline
$J^{n}$ & $-2Q_{J}S_{W}^{2}$ & $0$ & $\frac{-1+(1+2\sqrt{3}b
Q_{J})S_{W}^{2}}{\sqrt{3}\sqrt{1-(1+b ^{2})S_{W}^{2}}}$ & $\frac{%
-1+S_{W}^{2}}{\sqrt{3}\sqrt{1-(1+b ^{2})S_{W}^{2}}}$ \\ \hline
$U^{1}$ & $\frac{1}{2}-\frac{4}{3}S_{W}^{2}$ & $\frac{1}{2}$ & $\frac{-3+(3+5%
\sqrt{3}b )S_{W}^{2}}{6\sqrt{3}\sqrt{1-(1+b ^{2})S_{W}^{2}}}$ & $%
\frac{-1+(1-\sqrt{3}b )S_{W}^{2}}{2\sqrt{3}\sqrt{1-(1+b
^{2})S_{W}^{2}}}$ \\ \hline
$D^{1}$ & $-\frac{1}{2}+\frac{2}{3}S_{W}^{2}$ & $-\frac{1}{2}$ & $\frac{%
-3+(3-\sqrt{3}b )S_{W}^{2}}{6\sqrt{3}\sqrt{1-(1+b ^{2})S_{W}^{2}}}$
& $\frac{-1+(1+\sqrt{3}b )S_{W}^{2}}{2\sqrt{3}\sqrt{1-(1+b
^{2})S_{W}^{2}}}$ \\ \hline
$T^{1}$ & $-2Q_{T}S_{W}^{2}$ & $0$ & $\frac{1-(1-2\sqrt{3}b
Q_{T})S_{W}^{2}}{\sqrt{3}\sqrt{1-(1+b ^{2})S_{W}^{2}}}$ & $\frac{%
1-S_{W}^{2}}{\sqrt{3}\sqrt{1-(1+b ^{2})S_{W}^{2}}}$ \\ \hline
\end{tabular}%
\end{center}
\caption{Vector and axial vector couplings for the neutral weak currents $Z$ and $Z^{\prime }$ and for each fermion, with $i=1,2,3$, $n=2,3$, and the following electric charges for the singlets: $Q_{E}=(-1-\sqrt{3}b)/2$, $Q_J=(1+3\sqrt{3}b)/6$, and $Q_T=(1-3\sqrt{3}b)/6$.}
\label{EW-couplings}
\end{table}

\begin{table}[h]
\begin{center}
$%
\begin{tabular}{cccc}
\hline
Quantity & Experimental Values & Standard Model & 3-3-1 Model \\ \hline\hline
$\Gamma _{Z}$ $\left[ GeV\right] $ & 2.4952 $\pm $ 0.0023 & 2.4961 $\pm $
0.0010 & $\Gamma _{Z}^{SM}\left( 1+\delta _{Z}\right) $ \\ \hline
$\Gamma _{had}$ $\left[ GeV\right] $ & 1.7444 $\pm $ 0.0020 & 1.7426 $\pm $
0.0010 & $\Gamma _{had}^{SM}\left( 1+\delta _{had}\right) $ \\ \hline
$\Gamma _{\left( \ell ^{+}\ell ^{-}\right) }$ $MeV$ & 83.984 $\pm $ 0.086 & 
84.005 $\pm $ 0.015 & $\Gamma _{\left( \ell ^{+}\ell ^{-}\right)
}^{SM}\left( 1+\delta _{\ell }\right) $ \\ \hline
$\sigma _{had}$ $\left[ nb\right] $ & 41.541 $\pm $ 0.037 & 41.477 $\pm $
0.009 & $\sigma _{had}^{SM}\left( 1+\delta _{\sigma }\right) $ \\ \hline
$R_{e}$ & 20.804 $\pm $ 0.050 & 20.744 $\pm $ 0.011 & $R_{e}^{SM}\left(
1+\delta _{had}+\delta _{e}\right) $ \\ \hline
$R_{\mu }$ & 20.785 $\pm $ 0.033 & 20.744 $\pm $ 0.011 & $R_{\mu
}^{SM}\left( 1+\delta _{had}+\delta _{\mu }\right) $ \\ \hline
$R_{\tau }$ & 20.764 $\pm $ 0.045 & 20.789 $\pm $ 0.011 & $R_{\tau
}^{SM}\left( 1+\delta _{had}+\delta _{\tau }\right) $ \\ \hline
$R_{b}$ & 0.21638 $\pm $ 0.00066 & 0.21576 $\pm $ 0.00004 & $%
R_{b}^{SM}\left( 1+\delta _{b}-\delta _{had}\right) $ \\ \hline
$R_{c}$ & 0.1720 $\pm $ 0.0030 & 0.17227 $\pm $ 0.00004 & $R_{c}^{SM}\left(
1+\delta _{c}-\delta _{had}\right) $ \\ \hline
$A_{e}$ & 0.15138 $\pm $ 0.00216 & 0.1475 $\pm $ 0.0010 & $A_{e}^{SM}\left(
1+\delta A_{e}\right) $ \\ \hline
$A_{\mu }$ & 0.142 $\pm $ 0.015 & 0.1475 $\pm $ 0.0010 & $A_{\mu
}^{SM}\left( 1+\delta A_{\mu }\right) $ \\ \hline
$A_{\tau }$ & 0.136 $\pm $ 0.015 & 0.1475 $\pm $ 0.0010 & $A_{\tau
}^{SM}\left( 1+\delta A_{\tau }\right) $ \\ \hline
$A_{b}$ & 0.925 $\pm $ 0.020 & 0.9348 $\pm $ 0.0001 & $A_{b}^{SM}\left(
1+\delta A_{b}\right) $ \\ \hline
$A_{c}$ & 0.670 $\pm $ 0.026 & 0.6680 $\pm $ 0.0004 & $A_{c}^{SM}\left(
1+\delta A_{c}\right) $ \\ \hline
$A_{s}$ & 0.895 $\pm $ 0.091 & 0.9357 $\pm $ 0.0001 & $A_{s}^{SM}\left(
1+\delta A_{s}\right) $ \\ \hline
$A_{FB}^{\left( 0,e\right) }$ & 0.0145 $\pm $ 0.0025 & 0.01633 $\pm $ 0.00021
& $A_{FB}^{(0,e)SM}\left( 1+2\delta A_{e}\right) $ \\ \hline
$A_{FB}^{\left( 0,\mu \right) }$ & 0.0169 $\pm $ 0.0013 & 0.01633 $\pm $
0.00021 & $A_{FB}^{(0,\mu )SM}\left( 1+\delta A_{e}+\delta A_{\mu }\right) $
\\ \hline
$A_{FB}^{\left( 0,\tau \right) }$ & 0.0188 $\pm $ 0.0017 & 0.01633 $\pm $
0.00021 & $A_{FB}^{(0,\tau )SM}\left( 1+\delta A_{e}+\delta A_{\tau }\right) 
$ \\ \hline
$A_{FB}^{\left( 0,b\right) }$ & 0.0997 $\pm $ 0.0016 & 0.1034 $\pm $ 0.0007
& $A_{FB}^{(0,b)SM}\left( 1+\delta A_{e}+\delta A_{b}\right) $ \\ \hline
$A_{FB}^{\left( 0,c\right) }$ & 0.0706 $\pm $ 0.0035 & 0.0739 $\pm $ 0.0005
& $A_{FB}^{(0,c)SM}\left( 1+\delta A_{e}+\delta A_{c}\right) $ \\ \hline
$A_{FB}^{\left( 0,s\right) }$ & 0.0976 $\pm $ 0.0114 & 0.1035 $\pm $ 0.0007
& $A_{FB}^{(0,s)SM}\left( 1+\delta A_{e}+\delta A_{s}\right) $ \\ \hline
$Q_{W}(Cs)$ & $-$73.20 $\pm $ 0.35 & $-$73.23 $\pm $ 0.02 & $%
Q_{W}^{SM}\left( 1+\delta Q_{W}\right) $ \\ \hline
\end{tabular}%
\ \ $%
\end{center}
\caption{The parameters for experimental values, SM predictions and
331 corrections. The values are taken from ref. \protect\cite{one}}
\label{tab:observables}
\end{table}

\begin{table}[t]
\begin{tabular}{ll}
\hline
$\Gamma _{had}$ & $\Gamma _{\ell }$ \\ \hline\hline
1 &  \\ 
.39 & 1 \\ \hline
\end{tabular}%
\par
\begin{tabular}{lll}
\hline
$A_{e}$ & $A_{\mu }$ & $A_{\tau }$ \\ \hline\hline
1 &  &  \\ 
.038 & 1 &  \\ 
.033 & .007 & 1 \\ \hline
\end{tabular}%
\par
\begin{tabular}{llllll}
\hline
$R_{b}$ & $R_{c}$ & $A_{b}$ & $A_{c}$ & $A_{FB}^{(0,b)}$ & $A_{FB}^{(0,c)}$
\\ \hline\hline
1 &  &  &  &  &  \\ 
-.18 & 1 &  &  &  &  \\ 
-.08 & .04 & 1 &  &  &  \\ 
.04 & -.06 & .11 & 1 &  &  \\ 
-.10 & .04 & .06 & .01 & 1 &  \\ 
.07 & -.06 & -.02 & .04 & .15 & 1 \\ \hline
\end{tabular}%
\par
\begin{tabular}{llllllll}
\hline
$\Gamma _{Z}$ & $\sigma _{had}$ & $R_{e}$ & $R_{\mu }$ & $R_{\tau }$ & $%
A_{FB}^{(0,e)}$ & $A_{FB}^{(0,\mu )}$ & $A_{FB}^{(0,\tau )}$ \\ \hline\hline
1 &  &  &  &  &  &  &  \\ 
-.297 & 1 &  &  &  &  &  &  \\ 
-.011 & .105 & 1 &  &  &  &  &  \\ 
.008 & .131 & .069 & 1 &  &  &  &  \\ 
.006 & .092 & .046 & .069 & 1 &  &  &  \\ 
.007 & .001 & -.371 & .001 & .003 & 1 &  &  \\ 
.002 & .003 & .020 & .012 & .001 & -.024 & 1 &  \\ 
.001 & .002 & .013 & -.003 & .009 & -.020 & .046 & 1 \\ \hline
\end{tabular}%
\caption{The correlation coefficients for the Z-pole observables}
\label{tab:correlation}
\end{table}

\end{document}